\documentclass[aps,prb,superscriptaddress,twocolumn]{revtex4-1}

\usepackage{amsmath}
\usepackage{bbold}
\usepackage{amssymb}
\usepackage{amsthm}
\usepackage{graphicx}
\usepackage{graphics}
\usepackage{subfigure}
\usepackage{color}
\usepackage{bm}
\usepackage{hyperref}
\usepackage{rotating}

\DeclareSymbolFont{bbold}{U}{bbold}{m}{n}
\DeclareSymbolFontAlphabet{\mathbbold}{bbold}

\newcommand{\ket}[1]{|#1\rangle}

\newcommand{\be}{\begin{equation} }
\newcommand{\ee}{\end{equation} }
\newcommand{\ba}{\begin{eqnarray} }
\newcommand{\ea}{\end{eqnarray} }

\newcommand{\bpm}{\begin{pmatrix}}
\newcommand{\epm}{\end{pmatrix}}
\newcommand{\bmm}{\begin{matrix}}
\newcommand{\emm}{\end{matrix}}

\newcommand{\fops}[5]{F_{#1;#2 (#3)}^{\bar{#4}\bar{#5}}}
\newcommand{\fopb}[3]{F_{#1}^{#2,#3}}
\newcommand{\fop}[7]{F_{#1;#2 (#3)}^{(#4,#5)(#6,#7)}}

\begin{document}
\date{\today}

\begin{abstract}
Following an earlier construction of exactly soluble lattice models for abelian fractional topological insulators in two and three dimensions, we construct here an exactly soluble lattice model for a non-abelian $\nu=1$ quantum Hall state and a 
non-abelian topological insulator in two dimensions. We show that both models are topologically ordered, exhibiting
fractionalized charge, ground state degeneracy on the torus and protected edge modes. The models feature
non-abelian vortices which carry fractional electric charge in the quantum Hall case and spin in the
topological insulator case. We analyze the statistical properties of the excitations in detail and discuss the possibility of extending this construction to 3D non-abelian topological insulators.
\end{abstract}

\title{Exactly soluble lattice models for non-abelian states of matter in $2$ dimensions}
\author{Maciej Koch-Janusz}
\affiliation{Department of Condensed Matter Physics, Weizmann Institute of Science, Rehovot IL-76100, Israel}
\author{Michael Levin}
\affiliation{Condensed Matter Theory Center, Department of Physics, University of Maryland, College Park, Maryland 20742, USA}
\author{Ady Stern}
\affiliation{Department of Condensed Matter Physics, Weizmann Institute of Science, Rehovot IL-76100, Israel}
\maketitle

\section{Introduction}

The theoretical study of topological states of matter has greatly benefited from the discovery of exactly
soluble models whose ground states can be shown to be fractionalized topological states of matter.
Examples of such models are exactly soluble Hamiltonians for fractional quantum Hall states\cite{2007PhRvB..75g5318S},
the Toric Code\cite{AYu20032}, non-abelian string-net states \cite{2005PhRvB..71d5110L}, and models for
fractionalized topological insulators \cite{Levin:2011hq}. Topological states of matter are particularly
suitable for such studies, since their topological properties are protected against changes in the Hamiltonian
as long as the energy gap separating the ground state from the excited part of the spectrum does not close.

In an earlier work\cite{Levin:2011hq} we constructed exactly soluble models for fractionalized topological
insulators\cite{PhysRevLett.103.196803} in both two and three dimensions. Several properties make these
models ``fractionalized'': they have excitations that carry fractional charge, they have excitations that
follow fractional mutual statistics, and their ground state is degenerate when the system resides on a
topologically non-trivial manifold. Likewise, several properties make them topological insulators:
they are time reversal invariant and charge conserving, their bulk is gapped, and they have gapless surface modes that
are protected as long as time-reversal symmetry and charge conservation symmetry are not broken.
The key step in constructing these models was to couple non-interacting electrons to a solvable system of charged hard-core bosons that
realized a $\mathbb{Z}_k$ gauge theory. Even without interactions between the electrons themselves,
the interaction with the bosonic system was enough to drastically modify the low-energy physics,
which featured gapless surface and edge excitations with fractional charge and fermionic statistics,
as well as bulk excitations with fractional statistics.

In the models constructed in Ref. \onlinecite{Levin:2011hq} the mutual fractional statistics between
excitations was abelian. It is a natural to ask whether a similar construction could result in a \emph{non-abelian}
fractional topological insulator in 2 and 3 dimensions. Here we carry out the first steps of this program,
with a focus on the 2D case. By coupling the non-interacting electrons to a system of two flavors of
uncharged bosons we construct an exactly soluble model for a non-abelian $\nu=1$ quantum Hall state and
also a model for a 2D non-abelian fractional topological insulator. We also briefly discuss the generalization
to 3D non-abelian fractional topological insulators.

The bosonic Hamiltonian that underlies our construction is a variant of one of the generalized toric code models
of Kitaev.\cite{AYu20032} These models can realize any discrete gauge theory; here we consider the
simplest non-abelian gauge group $\mathbb{D}_3$. The resulting model has excitations that follow
non-abelian mutual statistics. Our strategy is to glue the excitations
of the bosonic model to electrons, and to put these composite particles in topologically nontrivial band structures
such as integer quantum Hall states and topological insulators.

The structure of the paper is as follows. Section \ref{summary} summarizes our results.
Sections \ref{Bosonicmodel1} and \ref{bosonicmodel2} introduce the bosonic model, first in an
abstract group theory language, based on Ref. \onlinecite{AYu20032}, and then a realization in a
system of bosons. Sections \ref{ribbons}, \ref{defects} introduce the notion of ribbon operators
and employ them to describe how topological defects are introduced to the bosonic system.
Section \ref{electrons} glues these defects to electrons, creating composite particles with
interesting properties. Section \ref{twodstates} introduces our main result: it defines a
lattice Hamiltonian that puts these composite particles into two topological
band structures -- a quantum Hall state and a 2D topological insulator state. Section \ref{lvsg}
explains the distinction between local and topological degrees of freedom and presents a
calculation of the ground state degeneracies in our model. Section \ref{statint} analyzes the statistical
interaction between different excitations of the model. We conclude and discuss the 3D case in section 
\ref{conclusions}. Three appendices give some technical details.

\begin{figure}[tb]
\centerline{
\includegraphics[width=0.9\columnwidth]{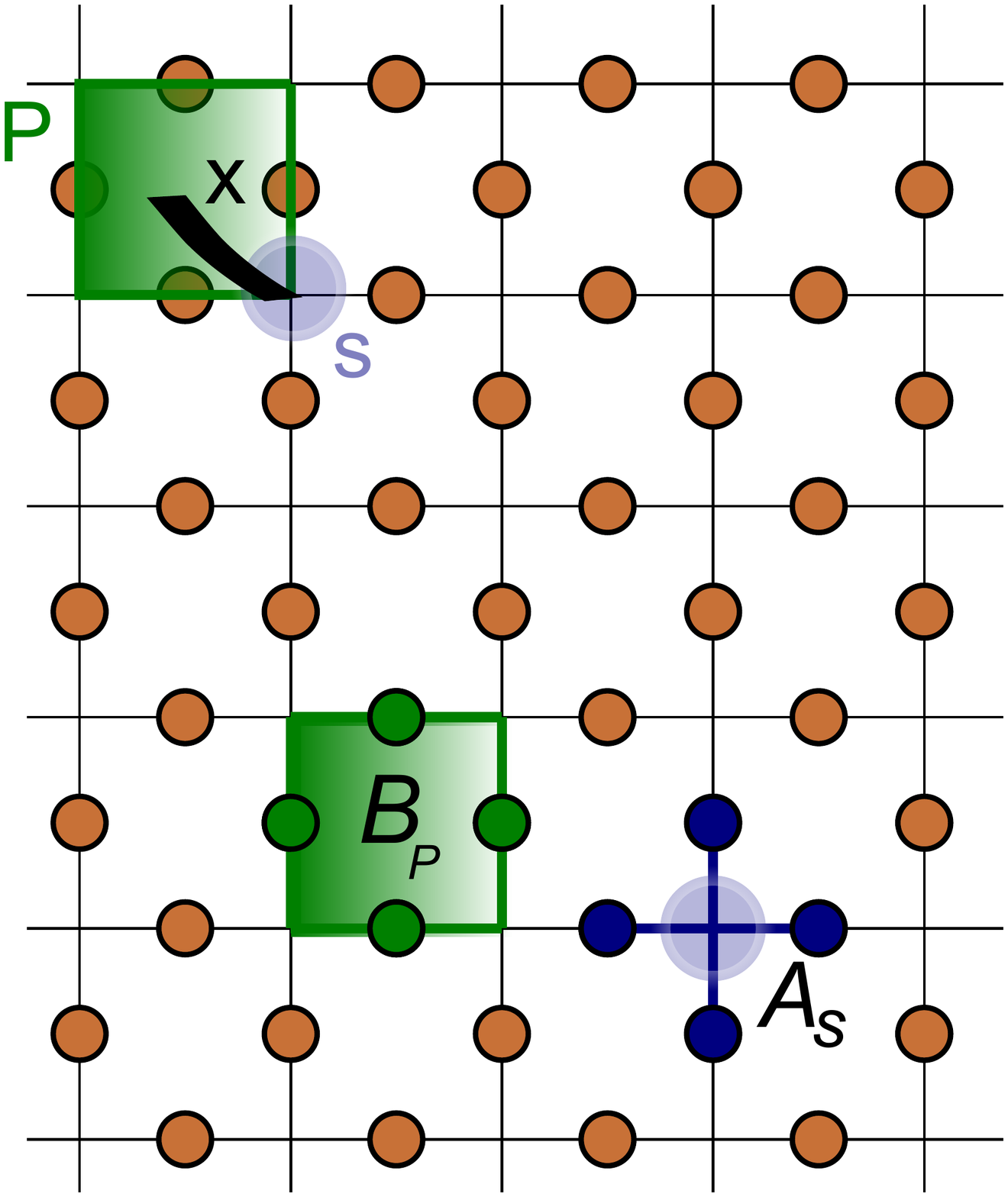}
}
\caption{The Hilbert spaces are on the links, which we have denoted with orange dots; we have suppressed link orientations in the picture. The operators $B$ act on the plaquettes, while the operators $A$ act the on stars. A combination of a plaquette $P$ and one of its vertices $s$ is  a site $x$.}
\label{rib5}
\end{figure}

\section{Summary of results\label {summary}}

This section is aimed at introducing the reader to our model and its properties, emphasizing the physical picture and leaving the details to the following sections.
\subsection{Constructing an exactly soluble model\label{summary-model}}
Our construction proceeds in a few steps: in the first step we create an exactly soluble lattice boson model whose
spectrum includes excitations with non-abelian mutual statistics.  In the second step we add electrons to the lattice
and design an electron-boson interaction which binds each electron to a carefully chosen abelian bosonic excitation. This binding 
creates a composite particle which is still a fermion with respect to other particles of its own kind,
but has nontrivial mutual statistics with other excitations of the bosonic system. In the third step we construct a
hopping term on the lattice that allows the fermionic composite to hop between lattice sites without exciting any
other degrees of freedom. Finally, we choose the hopping amplitudes so that the fermionic particles are placed in
one of the canonical 2D topological band structures: either a $\nu=1$ quantum Hall state or a 2D topological insulator
(we comment on the extension to non-abelian 3D topological insulator in the conclusion).

The bosonic model we construct in the first step is essentially a bosonic version of a generalized 'toric code' Kitaev Hamiltonian \cite{AYu20032}, based on a discrete non-abelian group $\mathbb{D}_3$. We consider a system of two flavors of bosons, $m$ and $n$, which live on the links $\langle ss' \rangle$ of a square lattice and we introduce a Hamiltonian composed out of two parts $H_B = H_1 + H_2$, both of which are projectors.

The first one, $H_1$, can be thought of as a 'charging term'; it assigns zero energy to certain preferred configurations $\{ n^n_{ss'},n^m_{ss'} \}$ of the bosonic occupation numbers (in the gauge theory picture they correspond to the absence of 'magnetic' fields) and it does so by coupling all the bosonic degrees of freedom on the links around a plaquette $P$. Other configurations are defined to be of positive energy. The spectrum of $H_1$ is discrete and highly degenerate.

The second term, $H_2$,  is the 'hopping' Hamiltonian, which makes the bosons hop between the neighboring links in a coordinated fashion. Likewise, $H_2$ has a discrete spectrum. A crucial aspect -- resulting in solubility -- is the fact that the two parts are mutually commuting: $[H_1,H_2]=0$. Thus the hopping Hamiltonian $H_2$ has matrix elements only between degenerate states of the charging Hamiltonian and it (partially) splits the degeneracy of the ground state. The residual degeneracy of the ground state depends on the topology and can be explicitly calculated: it is equal to one on the plane and is equal to the number of topologically distinct quasiparticle types in the theory -- i.e. 8 for the $\mathbb{D}_3$ case we consider -- when placed on the torus. The system is thus topologically ordered.

The second step of the construction involves introducing electronic degrees of freedom which live on the vertices $s$
of the lattice. We couple the electrons to the bosonic system by modifying the 'hopping' Hamiltonian ($H_2$ to $H_2^A$)
in such a way, that it is energetically preferential for the system to create one of the excitations of $H_2$ whenever
there is an electron present at a given vertex $s$. The precise excitation we choose is an abelian quasiparticle which we refer to as the ``$A$-charge.'' The word "charge" in the  term "A-charge" refers to topological, rather than electromagnetic, charge. The A-charge is electrically neutral, since all bosons in our model are neutral. The composite
excitation we create is then made up of an electron bound to an $A$-charge. This excitation has a unit electric charge, spin $1/2$. It has
fermionic self-statistics and nontrivial mutual statistics with respect to the other excitations of the bosonic system.

We then introduce a hopping term $H_{hop}$ for these composite excitations, which still commutes with all other terms
in the Hamiltonian, thus allowing the electron/A-charge particles to move around without dissociating. These composite
particles are low-energy excitations of our model and in many ways behave just like free electrons, so we can put them
in an electronic band structure of our choosing. Here we focus on two topologically nontrivial band structures:
(1) a $\nu = 1$ integer quantum Hall state and (2) a quantum spin Hall/topological insulator state.

\subsection{Properties\label{summary-properties}}

For either of the above band structures, the resulting model realizes a 2D non-abelian topological
phase: in the first case, the model realizes a non-abelian quantum Hall state with Hall conductance $\sigma_{H} = 1$,
while in the second case it realizes a non-abelian fractional topological insulator. In both
cases, the underlying topological order originates from the bosonic system.

To analyze this topological order, we recall that generalized Kitaev Hamiltonians, such as the one
used in our construction of the bosonic model, are in fact equivalent to discrete gauge theories --
in our case a $\mathbb{D}_3$ gauge theory. This allows us to identify the different excitations
present in the system, of which there are several types: the abelian $A$-charges and the non-abelian
$B$-charges, the $r$- and $\tau$-vortices and three types of dyons i.e. nontrivial
vortex/charge composites. Finally we have our electron/A-charge composites which carry a unit electric
charge and are semions with respect to the $\tau$-vortices.

The $r$- and $\tau$-vortices, as well as the $B$-charges and the dyons are non-abelian particles and therefore carry nonlocal, topological degrees of freedom. These degrees of freedom are insensitive to local perturbations and can only be accessed by braiding or fusing the non-abelian particles. We analyze in detail the topological braiding operation in a convenient basis of ribbon operators and explicitly show the connection to the associated
$\mathcal{R}$-matrix acting on the space of topological degrees of freedom.

The ribbon operators appearing in the model, which we denote by $F$, are a generalized form of the familiar string operators of the Toric Code, which create excitations of the system and whose support is strictly limited to links forming a \emph{ribbon} $\rho$, i.e. a narrow strip delimited by a path on the lattice and an adjacent path on a dual lattice as shown in Fig.(\ref{rib6}A). These operators commute with every term in the Hamiltonian except possibly the terms located at the ends $x_0, x_1$ of the ribbon (each end comprising one site and one plaquette). Well chosen ribbon operators can create excited eigenstates of the $H_1$ (associated with the plaquettes) or $H_2$ (associated with the vertices) parts of the Hamiltonian -- or both -- which correspond to vortex, charge and dyon type excitations, respectively. Furthermore, it is possible to derive many interesting properties of the excitations in the system, including braiding statistics and fusion just using the properties of the ribbon algebra (i.e. the commutation and other relations between various ribbon operators). We review the properties of the ribbon algebra in section V and explain in detail how to create an excited eigenstate using their linear combinations. We derive the braiding statistics for $\mathbb{D}_3$ and show that the physical picture is the one expected in the framework of a $\mathbb{D}_3$ gauge theory.

In the case of the non-abelian quantum Hall state, some of the quasiparticle excitations carry
fractional charge in addition to fractional statistics. In particular, we show that the
$\tau$-vortices have fractional charge $1/2$ (in units of $e$). This fractional charge is
important because it implies that these particles have smaller charge than any other quasiparticle
and therefore can be easily excited. We discuss a scheme for doing this and comment on how these
manipulations will effect non-abelian transformations on the state of the system.

In the case of the non-abelian fractional topological insulator, none of the excitations carry
fractional charge, but there are neutral excitations that carry the spin of an electron. 
This state also exhibits several other interesting properties
(in addition to non-abelian braiding statistics). First, it has a non-vanishing spin-Hall
conductivity: $\sigma_{sH} =1$ in units of $e/2\pi$.
Second, it has protected edge modes that cannot be gapped out without breaking time reversal or charge conservation symmetry. This combination justifies the name ``non-abelian 2D topological insulator.'' 


\section{The Kitaev Hamiltonian\label{Bosonicmodel1}}

In this section we review Kitaev's construction of a Hamiltonian that realizes a discrete gauge theory of a general group $G$. The construction was presented in Ref. \onlinecite{AYu20032} as well as Ref. \onlinecite{2008PhRvB..78k5421B}, and for some of the details we will refer the reader to these papers. In the next section we will translate this abstract construction to an explicit bosonic Hamiltonian.


The model is specified by a finite, discrete group $G$, in our case $G = \mathbb{D}_3$, and lives on the links of a square lattice (Fig.\ref{rib5}). The Hilbert space of the model is $\mathcal{H} = \bigotimes_{link} \mathcal{H}_{link}$ and $\mathcal{H}_{link} = \mathbb{C}[G]$, i.e. the complex Hilbert space for a single link is
spanned by the group elements of $G$. Each link comes with an orientation, whose change may be thought of as a basis change in which a basis vector $\ket{g}$ is mapped to its inverse $\ket{\bar{g}}$.

We now introduce the linear multiplication operators $L^\pm_g$ and projection operators $T^\pm_g$ labeled by the group elements, whose action on the basis states is:
\begin{gather}\label{t1} L^+_g\ket{h} = \ket{gh}, \mbox{\ \ \ } L^-_g\ket{h} = \ket{h\bar{g}}  \\
\label{t2} T^+_g\ket{h} = \delta_{g,h} \ket{h}, \mbox{\ \ \ } T^-_g\ket{h} = \delta_{\bar{g},h}\ket{h}. \end{gather}

The Hamiltonian is composed of two types of terms: the hopping terms, or 'gauge transformations' $A_s$ which live on the vertices and the charging terms, or 'Wilson loops' $B_p$ which live on the plaquettes of the lattice. The hopping term is:
\begin{equation}\label{as1} A_s = \frac{1}{|G|} \sum_g A_{s,g} = \frac{1}{|G|}\sum_{g\in G} L^{j_1}_g L^{j_2}_g L^{j_3}_g L^{j_4}_g, \end{equation}
where the labels $j_i$ refer to the orientations of the links incident on site $s$; $j_i$ is $+$ if the link is outgoing and $-$ if it is incoming.

The charging term on a plaquette $P$ with edges labeled $1, \ldots, 4$ is:
\begin{equation}\label{bp1} B_P = B_{P,e} =  \sum_{g_1g_2g_3g_4=e}T^{j_1}_{g_1}T^{j_2}_{g_2}T^{j_3}_{g_3}T^{j_4}_{g_4}, \end{equation}
where the links bordering the plaquette are taken in counterclockwise order; the orientations $j_i$ are positive if they agree with this order and negative otherwise. The subscript $e$ in $B_{P,e}$ denotes that the operator projects onto the states such that the product of group elements around the plaquette is the identity element $e$. Analogously we could define operators $B_{P,h}$ on each plaquette for an arbitrary $h\in \mathbb{D}_3$.

Both $A_s$ and $B_P$ operators are projectors, i.e.  their eigenvalues are either $0$ or $1$. The most important fact is, however, that all these operators commute with each other for all vertices $s,s'$ and plaquettes $P,P'$:
\begin{equation}\label{commute} [A_s,B_P] = [A_s,A_{s'}] = [B_P,B_{P'}]=0. \end{equation}
The Kitaev Hamiltonian is now given by:
\begin{equation}\label{kit1}  H_K = \sum_s (1-A_s) + \sum_P (1-B_P) \end{equation}

The crucial point about this Hamiltonian is that since all of its components commute, it can be diagonalized in the basis of the eigenstates of the $A_s$ and $B_P$ operators. Since all of its components are projectors the spectrum is discrete (in particular there is a gap between the ground state and the first excited state). The ground state has energy $E=0$ and corresponds to the eigenvalues of all $A_s$, $B_p$ operators being equal to $1$.
It is easy to see that the ground state of this Hamiltonian on the sphere is unique and is given by a totally symmetric superposition of all link configurations such that the product of elements around every plaquette is equal to identity. Its degeneracy on the torus is analyzed in section (\ref{lvsg}).

The excited states can be described in terms of localized particle excitations associated with \emph{sites} $x$,
which are formed by a plaquette $P$ and one of its vertices $s$ as shown in FIG.(\ref{rib5}). We will describe
the ribbon operators that create those excited states below, but we will here remark that the model can be shown
to be a lattice version of a discrete gauge theory \cite{deWildPropitius:1995hk} with gauge group $G$. Its
particles are labeled by pairs $(C),R$, where $(C)$ is one of the conjugacy classes of $G$ and $R$ is an
irreducible representation of the normalizer of a representative element chosen from $(C)$.


For our case of $G = \mathbb{D}_3$ the spectrum can be shown to contain the following particles:

\vspace{2mm}
\begin{tabular}{c|c||c|c}
$(C),R$ & type & $(C),R$ & type  \\
\hline
$(e), Id$ & vacuum & $(r),r1$ & dyon \\
$(e), A$ & abelian charge & $(r),r2$ & dyon \\
$(e), B$ & non-abelian charge & $(\tau), Id$ & vortex \\
$(r), Id$ & vortex & $(\tau),A$ & dyon
\end{tabular}
\vspace{3mm}

The 'charge' particles are excitations of the $A_s$ operators and are associated with sites. They always have bosonic self-statistics under a full braiding but can be either abelian (semion, in fact) or non-abelian with respect to the vortices. The vortices are excitations of the $B_P$ operators and live on the plaquettes, they are non-abelian with respect to vortices of the other type or even have non-abelian self-statistics ($\tau$-vortex) \cite{deWildPropitius:1995hk,FA198032}. Dyons can be thought of as mixed-type excitations. We will discuss the properties of the spectrum in greater detail when we describe the ribbon operators.

\section{The bosonic model}\label{bosonicmodel2}
As it stands, the Kitaev model is phrased in terms of abstract entities, with the Hilbert space being spanned by elements of group $\mathbb{D}_3$. We would like to introduce a more physical realization in terms of bosonic degrees of freedom on the links of the lattice. It turns out that due to the non-abelian nature of the gauge group we need at least \emph{two} flavors of bosons. (See Ref. \onlinecite{2005NJPh....7..187D} for a different approach).

Any element $g\in \mathbb{D}_3$ can be uniquely written as:
\begin{equation}\label{bosop0} g = \tau^n r^m, \end{equation} with $m=1,2,3$ and $n=1,2$. We introduce two flavors of electrically neutral bosonic degrees of freedom on the links of the lattice (for example, two types of atoms); the numbers $m,n$ will be the occupations of those bosonic states. In other words the bosonic Hilbert space on the links is spanned by the vectors $\ket{m,n}$ with $m=1,2,3$ and $n=1,2$. We introduce creation and annihilation operators $a$, $b$ for the m-bosons and n-bosons in the following fashion:
\begin{gather} \label{bosop1} [a_k,a^\dagger_l] = [b_k,b^\dagger_l] = \delta_{k,l} \\
\label{bosop2} [a_k,a_l] = [b_k,b_l] = [a_k,b_l] = [a_k,b^\dagger_l] = 0  \end{gather}

Let us examine the action of the operators $L^\pm_{\tau,r}$ on the bosonic states:
\[ L^+_r \tau^nr^m = r\tau^nr^m = \tau^n r^{-1^n}r^m = \tau^n r^{m+(-1)^n}, \]
from which it follows that:
\begin{equation}\label{bosop3} L^+_r\ket{m,n} = \ket{m+(-1)^n,n} \end{equation}
Analogously we can establish:
\begin{gather}\label{bosop4} L^-_r\ket{m,n} = \ket{m-1,n}, \mbox{\ \ \ } L^+_{\tau}\ket{m,n} = \ket{m,n+1} \\
\label{bosop5} L^-_{\tau}\ket{m,n} = \ket{-m,n+1}, \end{gather}
where all the arithmetic is understood to be mod 3 for the m-bosons and mod 2 for the n-bosons. In terms of $a$ and $b$ operators we have:
\begin{align} L^+_r &= [a^\dagger + (a)^2]bb^\dagger + [(a^\dagger)^2a + a]b^\dagger b,\\
L^-_r &= (a^\dagger)^2 + a; \mbox{\ \ \ \ }  L^+_{\tau} = b+b^\dagger,\\
 L^-_{\tau} &= [a^2(a^\dagger)^2+(a^\dagger)^2a + a^\dagger(a)^2](b+b^\dagger)  \end{align}
The $T^\pm_g$ operators can also be constructed in a similar fashion. For instance:
\begin{equation}\label{bosop7} T^\pm_e  = bb^\dagger[(a)^2(a^\dagger)^2], \mbox{\ \ \ \ \ } T^\pm_{\tau} = b^\dagger b[(a)^2(a^\dagger)^2] \end{equation}

By extension, we can also write bosonic versions of the charging and hopping terms in the Kitaev Hamiltonian, in fact any operator in that model, i.e. we have a fully bosonic description. Formally, for every operator $\hat{O}$ in Kitaev model we have a corresponding bosonic operator $B[\hat{O}]$. Let us denote by $H_B$ the bosonic Hamiltonian: $H_B = B[H_K]$.

\section{The ribbon operators for $\mathbb{D}_3$\label{ribbons}}

\begin{figure}[tb]
\centerline{
\includegraphics[width=0.9\columnwidth]{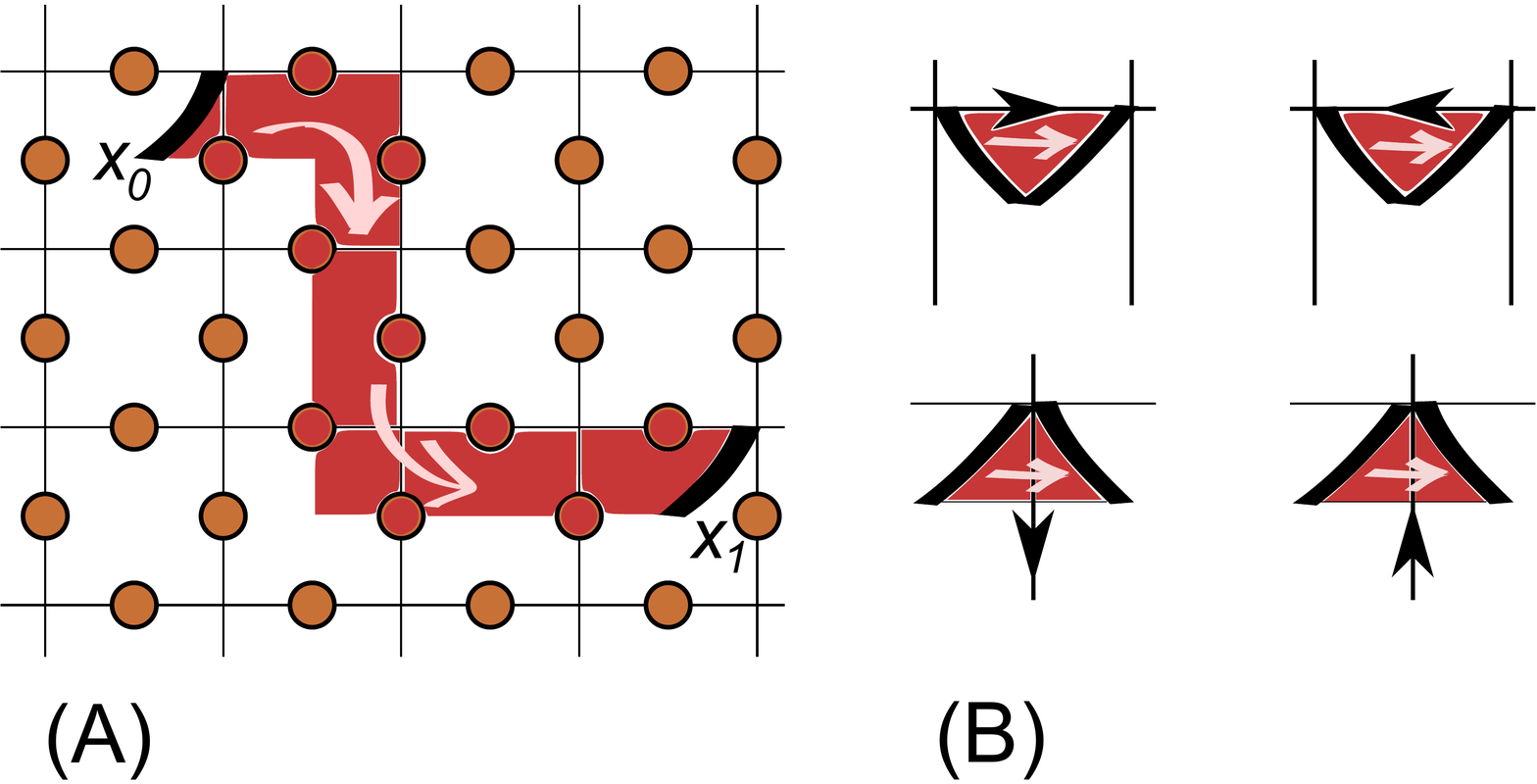}
}
\caption{(A) The ribbon operator between the sites $x_0$ and $x_1$; the arrow denotes the direction of the ribbon. The affected links (Hilbert spaces) have been colored red.
(B) Elementary ribbon operators. Every longer ribbon can be composed of elementary ones. The mutual orientations in the left row are defined as positive.}
\label{rib6}
\end{figure}

In the Kitaev model any n-particle excited state ($n \geq 2$) may be created by applying
\emph{ribbon operators} to the ground state. The ribbons are geometrical objects comprising links forming a directed narrow strip delimited by a path on the lattice and an adjacent path on a dual lattice. A ribbon operator operates only on links that are within the ribbon, as shown in Fig.(\ref{rib6}A). For an exhaustive treatment of ribbon algebra we refer to Ref.~\onlinecite{2008PhRvB..78k5421B}. Here we will review only those parts that are essential to our discussion. We will later use the terms \emph{ribbon} and \emph{ribbon operator} interchangeably where it does not cause confusion.

The algebra of ribbon operators along a directed ribbon $\rho$ with endpoints $x_0,x_1$ is
spanned by a set of operators $F_{\rho}^{h,g}$ with $h,g \in G$. The operators $F^{h,g}_{\rho}$
form a convenient and universal basis in which the algebraic manipulations are simple. To define
their action, it is useful to first consider \emph{elementary} ribbons of length $1$ or
\emph{triangles} from which any longer ribbon can be composed. Note that we have two distinct
types of triangles: those which contain a lattice link as one of their edges and those which
cross one of the links with their edge. We call them direct and dual, respectively; they are
depicted in Fig. (\ref{rib6} B). For a dual triangle, the ribbon operators are defined as
\begin{equation}\label{ribel1} F^{h,g}_{\rho} = \delta_{1,g}L^{j}_{h},\end{equation}
where the operator $L^{j}_{h}$ acts on the single lattice link which the triangle crosses.
Note that, though labeled by a pair of group elements, it only depends non-trivially on $h$.
This is done in order for the composition formula, eq.(\ref{ribel3}) below, to hold also for ribbons
of length one, which live on the direct or dual lattice but not both. For a direct triangle
\begin{equation}\label{ribel2} F^{h,g}_{\rho} = T^j_g, \end{equation}
where $T^j_g$ acts on the single link the triangle contains and analogously to the previous
case the operator depends non-trivially only on index $g$. The superscript $j=\pm$, depending
on the mutual orientation of ribbon and the edge, as shown in the lower and upper row of
Fig. (\ref{rib6} B). From those definitions it is clear that the operators on the dual triangles
are related to gauge transformations and the direct ones to projections.

Any operator on a longer ribbon $\rho = \rho_1 \cdot \rho_2$ can be decomposed into products
of operators living on shorter matching ribbons $\rho_1$ and $\rho_2$ in the following fashion:
\begin{equation}\label{ribel3} F^{h,g}_{\rho} = \sum_{c\in G} F_{\rho_1}^{h,c}F_{\rho_2}^{\bar{c}hc,\bar{c}g}. \end{equation}

A key property of the ribbon operators is that they commute with all the $A_s$ and $B_p$
operators (\ref{as1},\ref{bp1}) on the sites and plaquettes, except those at the two ends of
the ribbon. As a consequence, when a ribbon operator is applied to an eigenstate of the
Hamiltonian (\ref{kit1}), it creates local excitations at its ends. Moreover, when two ribbon
operators that start and end at the same points are applied to an eigenstate of the Hamiltonian,
the resulting states are the same, provided that there are no topological excitations within
the loop that they define. It is important to remark that the operators $F_{\rho}^{h,g}$ do
not create eigenstates, but their linear combinations do. We provide an explicit prescription
for constructing those combinations below.

The ribbon algebra is further characterized by the commutation properties of the ribbon operators. We will explicitly use only the following relation:
\begin{equation}\label{FopCom}\fopb{\rho_1}{h}{g}\fopb{\rho_2}{k}{l} = \fopb{\rho_2}{hk\bar{h}}{hl}\fopb{\rho_1}{h}{g},\end{equation}
which describes the exchange of two ribbons which both start at the same site, labeled by the pairs of group elements $(h,g)$ and $(k,l)$.

The eigenstates of the Kitaev Hamiltonian are created using superpositions of ribbons. Each excitation is labeled by a pair consisting of a conjugacy class $(C)$ and an irreducible representation $R$ of the normalizer of a representative element of $(C)$  (Recall that for a group element $g\in G$ the normalizer is the set of elements of $G$ which commute with $g$; it is a subgroup of $G$; normalizers of group elements in the same conjugacy class are isomorphic). Furthermore, depending on the values of $(C)$ and $R$, the ribbon has also indices $\bar{u} = (i,j)$, $\bar{v} = (i',j')$ describing the state at its two ends. These indices run over the range $i,i' = 1,\ldots,|C|$ and $j,j' = 1,\ldots,n_R$, with $n_R$ being the dimension of the irreducible representation $R$.

A general formula for a ribbon operator creating an eigenstate is then \cite{2008PhRvB..78k5421B}:
\begin{equation}\label{F} \fops{\rho}{R}{C}{u}{v} = \frac{n_R}{|N_C|} \sum_{n\in N_C} \bar{\Gamma}_R^{j,j'}(n) F_{\rho}^{\bar{c}_i,q_inq_{i'}}, \end{equation}
where $N_C$ is the normalizer of a representative element of the conjugacy class $C$ and $\Gamma_R(n)$ is the representation matrix associated to the element $n$ by the representation $R$. The overbar denotes complex conjugation.
The basis ribbon operators which enter the sum are those labeled by group elements in the appropriate conjugacy class and the coefficients are matrix elements of the appropriate representation of the normalizer elements.

We will explicitly construct some of the excitations for the case of $\mathbb{D}_3$, which will also elucidate the meaning of various notations in eq.(\ref{F}). To this end we need to introduce a certain fixed labeling of all the elements appearing in various algebraic objects taking part in our constructions, i.e. conjugacy classes, normalizers etc.
\begin{enumerate}
\item
 Conjugacy classes: $(e) = \{e\}$, $(r) = \{r,r^2\}$, $(\tau) = \{ \tau, \tau r, \tau r^2 \}$. We label the elements with $c_1$, $c_2$ etc. in the order they appear in each class.
\item
 We choose a representative $r_C$ of each conjugacy class: the first element, i.e. $c_1$, in each class.
\item
 For each class we construct normalizers $N_{r_C}$ of $r_C$. Different choice of the representative $r_C$ in principle produces different normalizers but they are all isomorphic. We set $N_e = \mathbb{D}_3$, $N_r = \{e,r,r^2\} \simeq \mathbb{Z}_3$, $N_{\tau} = \{e,\tau\} \simeq \mathbb{Z}_2$
\item
 We construct the cosets $Q_C = \mathbb{D}_3/N_C $:
 $Q_e = \{e\}$, $Q_r = \{ e,\tau \}$, $Q_{\tau} = \{ e,r,r^2\}$ and label their elements as $q_1$, $q_2$, $\ldots$ in order they appear.
\item
 Any $g\in C \subseteq \mathbb{D}_3 $ can be uniquely decomposed into $g=q_in$ with $q_i \in Q_C$ and $n\in N_C$

 \begin{tabular}{c|c|c||c|c|c}
 $g=q_in$ & i(g) & n(g) & $g=q_in$ & i(g) & n(g) \\
 \hline
 $e=ee$ & $1$ & $e$ & $\tau = e\tau$ & $1$ & $\tau$ \\
 $r=er$ & $1$ & $r$ & $\tau r = r^2\tau$ & $3$ & $\tau$ \\
 $r^2=er^2$ & $1$ & $r^2$ & $\tau r^2 = r\tau$ & $2$ & $\tau$
 \end{tabular}
 
This defines $i(g)$ and $n(g)$.
\item
 The final step is to consider the irreducible representations of the normalizers. For the conjugacy class $(e)$ we have $N_e = \mathbb{D}_3$, whose representations $\Gamma$ are: identity, $\Gamma_{Id}(g) = 1$ for all $g$; alternating, $\Gamma_A(\tau^nr^m)=(-1)^n$ and a two-dimensional representation defined by $\Gamma_{B}(r) = \left( \begin{array}{cc} e^{2\pi i/3} & 0 \\ 0 & e^{-2\pi i/3} \end{array} \right) $ and $\Gamma_B(\tau) = \left( \begin{array}{cc}0&1\\1&0 \end{array} \right)$.

 For the conjugacy class $(r)$ we have $N_r \simeq \mathbb{Z}_3$ with three 1-dimensional irreducible representations: identity $\Gamma_{(r)Id}$, and two others defined by $\Gamma_{r1}(r) = e^{2\pi i/3}$ and $\Gamma_{r2}(r) = e^{-2\pi i /3}$, respectively.

  For the $(\tau)$ class, $N_{\tau}\simeq \mathbb{Z}_2$ and we have only two irreducible representations: identity $\Gamma_{(\tau)Id}$ and an alternating one given by $\Gamma_{(\tau)A}(\tau) = -1$.
\end{enumerate}

Now we can write out explicitly the operators creating some of the interesting  excitations in our system using formula (\ref{F}). Let us begin with pure charges corresponding to the trivial conjugacy class. Vacuum is a special type:
\[ \fop{\rho}{Id}{e}{1}{1}{1}{1} = \frac{1}{6} \left( \fopb{\rho}{e}{e} + \fopb{\rho}{e}{r}+\fopb{\rho}{e}{r^2}+\fopb{\rho}{e}{\tau}+\fopb{\rho}{e}{\tau r}+\fopb{\rho}{e}{\tau r^2} \right) \]
The abelian A-charges and non-abelian B-charges:
\[ \fop{\rho}{A}{e}{1}{1}{1}{1} = \frac{1}{6}\left( \fopb{\rho}{e}{e} + \fopb{\rho}{e}{r} + \fopb{\rho}{e}{r^2} - \fopb{\rho}{e}{\tau} - \fopb{\rho}{e}{\tau r} -\fopb{\rho}{e}{\tau r^2} \right)  \]
\begin{align*} \fop{\rho}{B}{e}{1}{1}{1}{1} &= \frac{1}{3}\left( \fopb{\rho}{e}{e} + e^{-2\pi i/3}\fopb{\rho}{e}{r} + e^{2\pi i/3}\fopb{\rho}{e}{r^2} \right) \\
 \fop{\rho}{B}{e}{1}{2}{1}{2} &= \frac{1}{3}\left( \fopb{\rho}{e}{e} + e^{2\pi i/3}\fopb{\rho}{e}{r} + e^{-2\pi i/3}\fopb{\rho}{e}{r^2} \right) \\
 \fop{\rho}{B}{e}{1}{1}{1}{2} &= \frac{1}{3}\left( \fopb{\rho}{e}{\tau} + e^{2\pi i/3}\fopb{\rho}{e}{\tau r} + e^{-2\pi i/3}\fopb{\rho}{e}{\tau r^2} \right) \\
 \fop{\rho}{B}{e}{1}{2}{1}{1} &= \frac{1}{3}\left( \fopb{\rho}{e}{\tau} + e^{-2\pi i/3}\fopb{\rho}{e}{\tau r} + e^{2\pi i/3}\fopb{\rho}{e}{\tau r^2} \right) \end{align*}
Note that unlike the abelian A-charge, the non-abelian B-charge has two internal states. The indices $j,j'$ run from $1$ to $2$, so in total we have four operators for different states at the two ends of the ribbon. Analogously for the $r-vortex$:
\begin{align*} \fop{\rho}{Id}{r}{1}{1}{1}{1}&=\frac{1}{3}\left(\fopb{\rho}{r^2}{e} + \fopb{\rho}{r^2}{r}+\fopb{\rho}{r^2}{r^2} \right)\\
 \fop{\rho}{Id}{r}{2}{1}{2}{1}&=\frac{1}{3}\left(\fopb{\rho}{r}{e} + \fopb{\rho}{r}{r}+\fopb{\rho}{r}{r^2} \right)\\
 \fop{\rho}{Id}{r}{1}{1}{2}{1}&=\frac{1}{3}\left(\fopb{\rho}{r^2}{\tau} + \fopb{\rho}{r^2}{\tau r}+\fopb{\rho}{r^2}{\tau r^2} \right)\\
 \fop{\rho}{Id}{r}{2}{1}{1}{1}&=\frac{1}{3}\left(\fopb{\rho}{r}{\tau} + \fopb{\rho}{r}{\tau r}+\fopb{\rho}{r}{\tau r^2} \right)\end{align*}

It is equally simple to construct operators for all other excitations. 

\begin{figure}[tb]
\centerline{
\includegraphics[width=0.9\columnwidth]{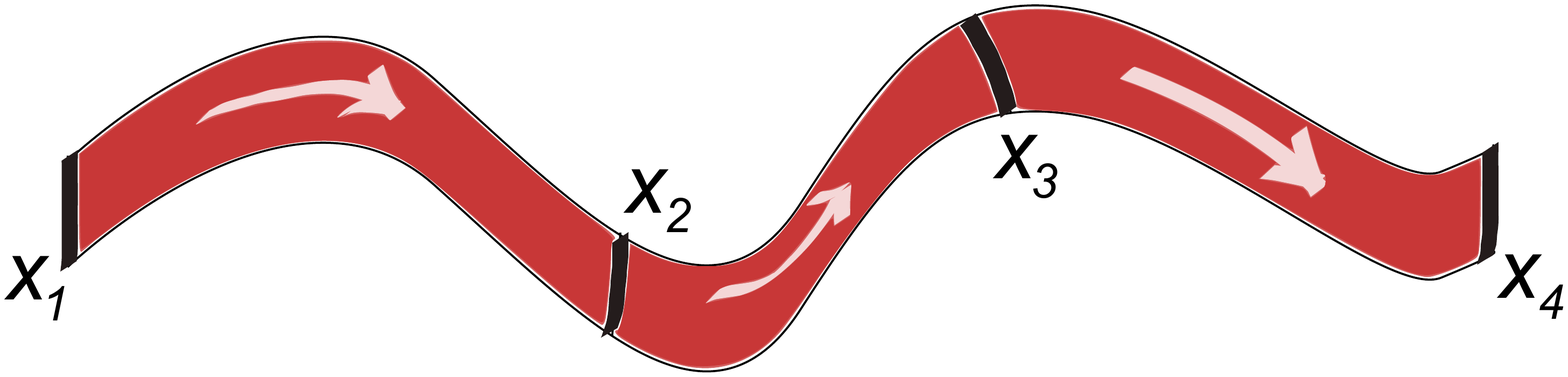}
}
\caption{Ribbon operators creating excitations at sites $x_1,\ldots,x_4$}
\label{rib2}
\end{figure}

\section{Introducing particles into the ground state\label{defects}}

The ground state $\ket{gs}_{H_k}$ of the Kitaev Hamiltonian obviously does not contain any of the excitations described above. However, we can introduce a modified Hamiltonian $H'$ such that its ground state does contain a number of such particles. To this end, let us first introduce operators $A_s^{R,k}$ with $k=1,\ldots, n_R$ as a generalization of the $A_s$:
\begin{equation}\label{asr1} A_s^{R,k} = \frac{1}{|G^k_R|}\sum_{g\in G} \Gamma_R^{kk}(g) L^{j_1}_g L^{j_2}_g L^{j_3}_g L^{j_4}_g, \end{equation}
where $|G^k_R|$ is the number of elements $g\in G$ such that $\Gamma_R^{kk}(g) \not= 0$.
Note that the choice $R = Id$ reproduces the operator $A_s$. The operators $A_s^{R,k}$ are projectors, even in the case of $R$  being a higher-dimensional representation.
Also, they commute with each other as well as with the $B_P$ operators:
\begin{equation}\label{modcom} [A_s^{R,k}, B_P] = [A_s^{R,k},A_{s'}^{R,k}] =0  \end{equation}

Labeling those operators with the names of the \emph{excitations} of the Kitaev Hamiltonian is justified by the following observation: imagine creating a pair of $A$-type charges from Kitaev vacuum using a ribbon operator with endpoints $s_0,s_1$: $F_{s_0,s_1}^A\ket{g.s.}_{H_K}$, we know this is an excited state of $H_k$ since it is annihilated by the $s_1$ term of the Hamiltonian:  $A_{s_1} F_{s_0,s_1}^A\ket{g.s.}_{H_K} = 0$. However it is also true that this state is invariant under the action of the modified operator: $A_{s_1}^A F_{s_0,s_1}^A\ket{g.s.}_{H_K} = F_{s_0,s_1}^A\ket{g.s.}_{H_K}$, i.e. the ground state of a Hamiltonian $H'$ with $A_s^A$ operators in place of $A_s$ looks like an excited state of $H_k$. We shall include a symmetric combination in the Hamiltonian:
\begin{equation}\label{assym} A_s^{B} = A_s^{B,1} + A_s^{B,2}. \end{equation}
A brief discussion of the $A_s^{R,k}$ operators may be found in Appendix C.

We may also introduce modified flux operators, which project onto states such that the product of the elements around a plaquette is in the conjugacy class $(C)$:
\begin{equation}\label{bpc1} B_P^{(C)} = \sum_{g_1\cdots g_4 \in (C)} T_{g_1}^{j_1}T_{g_2}^{j_2}T_{g_3}^{j_3}T_{g_4}^{j_4}. \end{equation}
As a special case we have $B_P = B_{P,e} = B_P^{(e)}$. Note that including the projections $B_{P,h}$ onto individual group elements $h\in\mathbb{D}_3$ in the Hamiltonian (apart from the identity $e$ which is a one-element conjugacy class by itself) conflicts with its exact solubility, since such projections would not commute with the $A_s^R$ operators -- their action only preserves the conjugacy class.

\section{Coupling to the electronic degrees of freedom\label{electrons}}

Let us now introduce electronic degrees of the freedom on the \emph{vertices} of the lattice. We denote the creation/annihilation operators for the electron at site $s$  and spin $\sigma$ by $c^{\dagger}_{s\sigma}$, $c_{s\sigma}$. Let $n_{s\sigma}$ be the number of electrons of spin $\sigma$ at that site, and $n_s=\sum_\sigma n_{s\sigma}$ the total number of electrons on the site.
We will now glue an A-charge to every electron, by modifying the Hamiltonian to
\begin{equation}\label{hbos2} H_1 = \sum_s (1-A^{A,n}_s) + \sum_P (1-B_P) - \mu\sum_s n_s \end{equation}
with $\mu$ being the chemical potential for the electrons, and
\begin{equation}\label{asr2} A^{A,n}_s  =  \frac{1}{|G|}\sum_{g\in G} \left(\Gamma_A(g) \right)^{n_{s}} L^{j_1}_g L^{j_2}_g L^{j_3}_g L^{j_4}_g\end{equation}
Since $\Gamma_A$ is the 1-dimensional alternating representation it only assumes values $\pm1$. Therefore, in the presence of an electron at site $s$ this factor stays unchanged and $A^{A,n=1}_s = \Gamma_A(g) = A^{A}_s $. In contrast, in the absence of an electron, or when the site has two electrons, we have $\Gamma_A(g)^0 = 1$ for all $g\in G$ and therefore $A^{A,n=0}_s = A^{V}_s = A_s$. Thus we now have in the ground state an $A-charge$ at vertex $s$ whenever there is one electron there, i.e. we  bind an $A$-charge to each electron. Two electrons on the same site, with opposite spin directions, are glued to two $A$-charges, which fuse to the vacuum.

So far the model is static. To introduce dynamics, we add a hopping term for the electrons modified in such a way that the electrons hop together with their associated $A$-charges. Since the operator that hops an $A$-charge from site $s$ to $s'$ is the elementary ribbon  $F^A_{ss'}$, the operator
$c^{\dagger}_{s'}F^{A}_{ss'}c_{s}$ will move the electron/A-charge composite in one piece. To make it more general we also introduce  arbitrary hopping amplitudes $t_{ss'}$ and add to the Hamiltonian the hopping term
\begin{equation}\label{hhop1} H_{hop} =  - \sum_{<ss'>}\left( t_{ss'}c^{\dagger}_{s'}F^{A}_{ss'}c_{s} + h.c.  \right), \end{equation}
where the summation in the hopping term is over all links. It is easy to check that $H_{hop}$ commutes with all the other operators of the Hamiltonian.

\section{Non-abelian $\nu=1$ quantum Hall state and 2D non-abelian topological insulator\label{twodstates}}

At this point we have a Hamiltonian which in the low-energy subspace looks much like a tight-binding model
of usual electrons because the $A$-charges bound to the electrons are bosons with respect to each other.
We can therefore choose the hopping amplitudes so that the composite fermionic particles are put into any
band structure we like. Here we consider two topologically nontrivial band structures:
(1) a $\nu = 1$ integer quantum Hall state, and (2) a 2D quantum spin Hall state/topological insulator.

We begin with case (1) where the composite fermions are put into a $\nu = 1$ integer quantum Hall state.
In this case, the model realizes a non-abelian quantum Hall state with Hall conductance $\sigma_H = 1$.
An interesting feature of this state is that it supports a fractionally charged non-abelian vortex --
specifically the $\tau$-vortex. To see this, we recall that the $\tau$-vortex is a semion with respect
to the $A$-charge, i.e. the $A$-charge obtains a $\pi$ phase upon winding around the vortex. It follows that
the $A$-charge-electron composite also obtains a $\pi$ phase upon winding around the vortex. The crucial
observation is that this phase is precisely the same as an electron in a $\nu=1$ integer quantum Hall
state would acquire upon winding a threaded electromagnetic flux of half of a flux-quantum (i.e. a $\pi$-flux).
Thus the introduction of a $\tau$-vortex will elicit the same response in our composite quantum Hall
system as the introduction of a $\pi$-flux in a $\nu = 1$ integer quantum Hall state of electrons.
This response is a localized screening charge of $\pm 1/2$, and thus the non-abelian $\tau$-vortices
carry fractional charge $1/2$ (in units of $e$).

An important consequence of this observation is that the $\tau$-vortices are the excitations with
the minimal charge. This means that the system will preferentially develop such excitations whenever it
is forced to minimize the Coulomb energy associated with charge non-uniformity. This gives us natural
ways to introduce vortices. As an example, consider adding the following  
term into the Hamiltonian:
\begin{equation}\label{himp1} H_{imp} =  V\left(\sum_{s\in\partial P} n_s - \frac{1}{2} \right) ^2, \end{equation}
where the sum is over the corner vertices of some plaquette $P$. If we make the parameter $V$ very
large the system will prefer an excess $1/2$ electric charge in that region in order to minimize the
energy. Thus the Hamiltonian term in eq.(\ref{himp1}) will force a $\tau$-vortex at plaquette $P$.
Physically this term describes an impurity binding an isolated $\tau$-vortex. We can now imagine having
a number of such well-separated impurities and tuning their respective parameters $V$ in the Hamiltonian
to drag vortices from one plaquette to another. In this way we can implement braiding of the non-abelian
vortices and therefore the state of the system would undergo non-abelian transformations.

We now consider case (2) where the composite fermions (the electron-A-charge composites) are put into a 2D quantum spin Hall state.
More specifically, we consider a band structure where the spin $\uparrow$ and $\downarrow$ 
fermions form $\nu=1$ quantum Hall states with opposite chiralities. This state has a vanishing
electric Hall conductance, and a spin-Hall conductance of $\sigma_{sH} = 1$ in units of
$e/2\pi$.

Unlike the quantum Hall state discussed above, this spin-Hall state does not support any excitations
with fractional charge. 
On the other hand, the state does support excitations that are electrically neutral but carry the same $s^z$ spin as an electron. 
An example of such an excitation is the $\tau$-vortex: following the same reasoning as in case (1) discussed above, we can see
that the $\tau$-vortex will bind half a fermion of one spin direction and half a hole of the other spin direction. 

In addition, this system has protected edge modes which cannot be gapped out unless time reversal
or charge conservation symmetry is broken (explicitly or spontaneously). To see this, note that
the minimal quasiparticle charge in this system is $e^* = 1$ (in units of $e$). Thus, the ratio of
the spin-Hall conductance to the elementary charge is $\sigma_{sH}/e^* = 1$, an odd number.
Applying the general flux insertion argument of Ref. \onlinecite{PhysRevLett.103.196803},
we conclude that this system must have protected edge modes.

\section{Non-abelian statistics: structure of the Hilbert space\label{lvsg}}

We now analyze in greater detail the non-abelian statistics in our model and in particular
the Hilbert space structure associated with the non-abelian statistics. Given that this non-abelian structure is
completely dictated by the underlying bosonic system, we will neglect the electronic
degrees of freedom and focus entirely on the bosonic model. We note that this section is largely a pedagogical review of the analysis in Ref. \onlinecite{AYu20032}.

Let us fix $n$ arbitrary sites $x_1,\ldots,x_n$ and let us consider the projected Hilbert space 
$\mathcal{L}_n(x_1,\ldots,x_n)$, i.e. the Hilbert space of $n$ particles restricted to those fixed sites. 
Since every $n$-particle state can be obtained using $n-1$ ribbon operators in the fashion shown in Fig.\ref{rib2} 
and the space of ribbons is spanned by $|G|^2$ operators labeled by the pairs $(g,h)\in\mathbb{D}_3$, the 
dimension of this space is $|G|^{2(n-1)}$.
It is natural to classify this space first by particle types, and then to distinguish between local and 
topological degrees of freedom of any given combination of particles. This distinction can be understood 
by considering the effects of \emph{local} operators whose spatial support is limited to the region around 
one of the points $x_i$ and which preserve the space of n-particle excitations and its orthogonal complement. 
It can be shown that operators that do not change particle types can be written as $D_{(h,g)}=B_{h}A_{g}$, 
with the operators $A_{g}$,  $B_{h}$ acting around one of the sites $x_i$ and defined by eqs.(\ref{as1},\ref{bp1}) 
and the discussion below them. These operators do not create new excitations at other sites nor do they connect 
any of the sites in $x_1,\ldots,x_n$, which are assumed to be far apart. We denote by $\mathcal{D}$ the algebra they generate.


We can then consider $\mathcal{D}(x_1),\ldots,\mathcal{D}(x_n)$ i.e. the algebras of local operators acting around sites $x_1, \ldots, x_n$ (these algebras are all isomorphic to $\mathcal{D}$) and the algebra $\mathcal{P}_n$ generated by all of them.
 The center of each of the algebras $\mathcal{D}(x_i)$ is the set of projectors on different particle types, since local operators cannot change the particle type at the point where they operate.

Under the action of the local perturbations the $n$-particle Hilbert space splits in the following fashion:
\begin{equation}\label{struct1} \mathcal{L}_n = \bigoplus_{d_1,\ldots,d_n} \mathcal{L}_{d_1,\ldots,d_n}, \end{equation}
with $d_i = R,(C)$ denoting the type of particle. Also the algebra $\mathcal{P}_n$ splits similarly (i.e. all of the operators have a block-diagonal form), since local perturbations cannot change particle types. Furthermore the space $\mathcal{L}_{d_1,\ldots,d_n}$ decomposes under the perturbations from $\mathcal{P}_{d_1,\ldots,d_n}$:
\begin{equation}\label{struct2} \mathcal{L}_{d_1,\ldots,d_n} = \mathcal{K}_{d_1}\otimes\ldots\otimes\mathcal{K}_{d_n}\otimes\mathcal{M}_{d_1,\ldots,d_n} \end{equation}
The spaces $\mathcal{K}_{d_i}$ are the local degrees of freedom of the $i$-th particle, which can be changed by the local perturbations. The space $\mathcal{M}_{d_1,\ldots,d_n}$ however does not have a tensor product structure and is insensitive to local perturbations. This is the space of the non-local degrees of freedom.

The space $\mathcal{M}_{d_1,\ldots,d_n}$, also called \emph{the protected space}, undergoes unitary transformations when particles are braided and grows or shrinks when they are fused/created out of vacuum. Thus it is only influenced by the topological operations. The dimensionality of this space is determined by the fusion rules. The fusion rules for the quasi-particles of the $\mathbb{D}_3$ group are given in Appendix A.

\begin{figure}[tb]
\centerline{
\includegraphics[width=0.9\columnwidth]{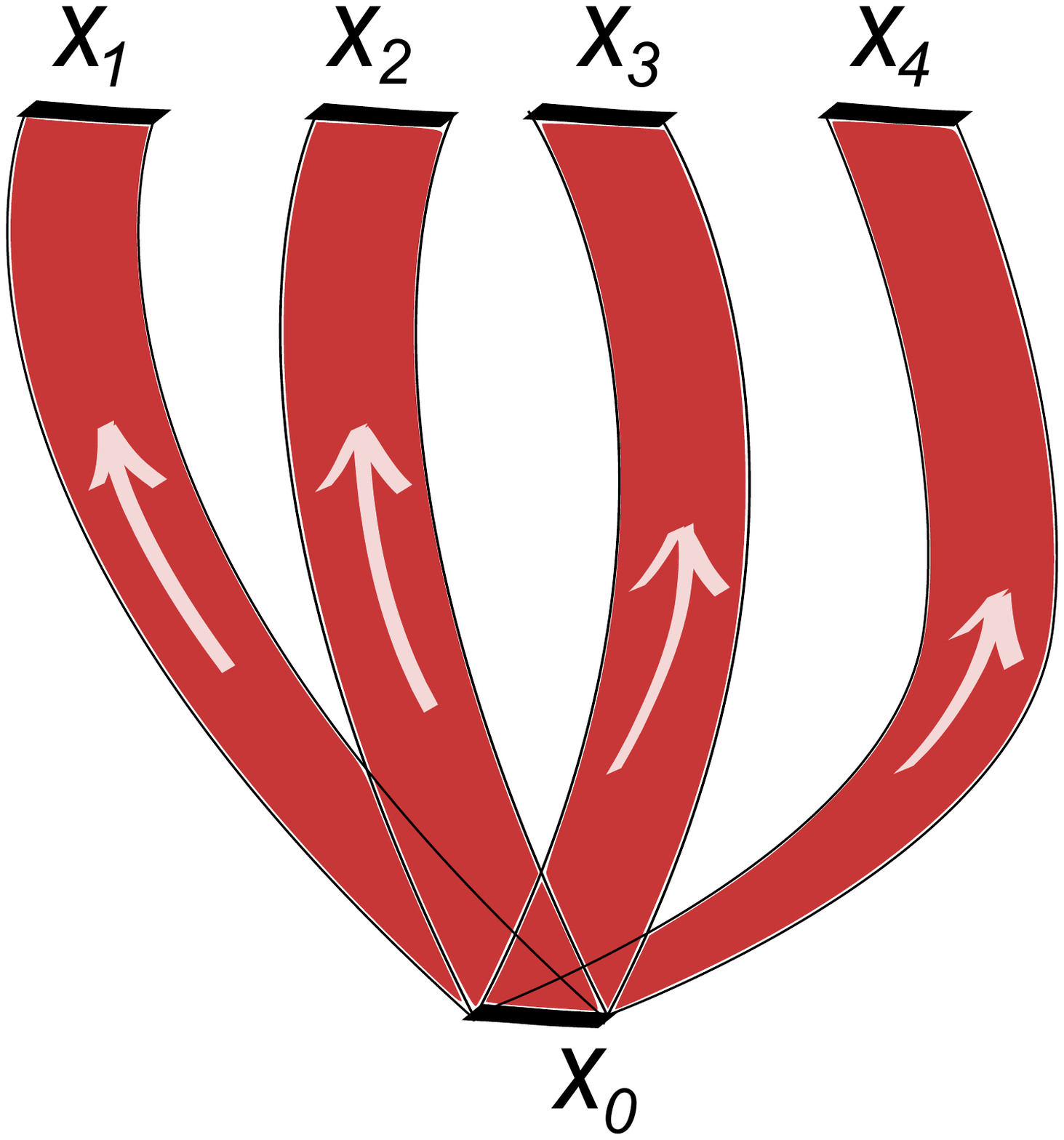}
}
\caption{Ribbon operators creating excitations at sites $x_1,\ldots,x_4$ anchored at an auxiliary site $x_0$}
\label{rib1}
\end{figure}

The dimensionality of the protected space is smaller than the full dimension of the Hilbert space, since not all degrees of freedom are topological. This difference may be elucidated by the following construction:
consider the $n$-particle space as a subspace of the $(n+1)$-particle space $\mathcal{L}(x_0,x_1,\ldots,x_n)$ spanned by only those ribbon operators which commute with all local perturbations from $\mathcal{D}(x_0)$. Let us arrange the ribbons as in Fig.(\ref{rib1}). The commutation condition at $x_0$ then implies that at site $x_0$ all the ribbons fuse to vacuum, since this is the only ribbon operator which commutes with all local perturbations. In other words, physically there is nothing at site $x_0$. However, every ribbon $\alpha$ has a pair of indices $\bar{u}_{\alpha}=(i_{\alpha},j_{\alpha})$ associated with $x_0$. The vacuum condition means they are all contracted in such a way that when an operator from $\mathcal{D}({x_0})$ is applied and commuted through them, changing all of them in the process, the contraction remains the same. More concretely, the creation of a physical state with particles $d_1,\ldots,d_n$ at sites $x_1,\ldots,x_n$ in a manner shown in Fig. (\ref{rib1}) involves the application to the ground state of an operator of the form:
\begin{equation}\label{W1}\sum_{\bar{u}_1\ldots \bar{u}_n} W^{(k)}_{\bar{u}_1\ldots \bar{u}_n}(d_1,\ldots,d_n)F^{\bar{u}_1,\bar{v}_1}_{\rho_1,d_1}\cdots F^{\bar{u}_n,\bar{v}_n}_{\rho_n,d_n}, \end{equation}
which we will symbolically write as $W^{(k)}F_1\cdots F_n$.

The vacuum condition can be then phrased as:
\begin{equation}\label{W2} D_{(h,g)}W^{(k)}F_1\cdots F_n = W^{(k)}F_1\cdots F_n D_{(h,g)},\end{equation}
for all the $D_{(h,g)}$ perturbations from $\mathcal{D}(x_0)$.


The index $k$ labels different tensors with this property. The number of different possible tensors $W(d_1,\ldots,d_n)$ is the number of distinct ways in which particles $d_1,\ldots,d_n$ can be fused to vacuum and so precisely the dimension of the protected space $\mathcal{M}_{d_1,\ldots,d_n}$. Knowing the fusion rules it is therefore possible to  calculate this number explicitly using Bratelli diagrams. In other words different $W(d_1,\dots,d_n)$ tensors label states that may be distinguished only by non-local measurements.


 The local operators from $\mathcal{D}(x_1),\ldots,\mathcal{D}(x_n)$ acting on the physical sites can change the  indices at those sites, but they cannot influence the indices at $x_0$. Thus in a certain sense the indices at $x_0$ are 'topological'. Let us explain what we mean by that. In addition to the local operators there are also \emph{topological operators}. These operators commute with all the elements of $\mathcal{D}(x_1)\otimes\ldots\otimes\mathcal{D}(x_n)$. When applied within the $n$-particle subspace, they transform the system from one $W$ tensor to another. They do so by changing some ribbon indices at the point $x_0$. However, they should not be regarded as local operators that belong to $\mathcal{D}(x_0)$. Operators that belong to $\mathcal{D}(x_0)$ act on all ribbons that originate at $x_0$, and hence change the indices of all these ribbons. In contrast, topological operators can change the individual indices of one ribbon at $x_0$ without changing the indices of any other ribbon. The set can be spanned by operators $D^{\alpha}_{(h,g)}$ which \emph{only} act on ribbon $\alpha$ in the same way an operator $D_{(h,g)}$ would, but leave the other ribbons untouched.
   It is thus possible for the topological operators to change non-trivially (i.e. in a way that cannot be brought to the original by just relabeling the summation indices) the coefficients of the contraction (\ref{W1}) at site $x_0$, or equivalently to map between different tensors $W^{(k)}(d_1,\ldots,d_n)$:
\begin{equation}\label{W3} D^{\alpha}_{(h,g)}W^{(k)} = U_{kk'}W^{(k')} . \end{equation}
This physically corresponds to moving between different non-local states of the particles.

 For $n=2$ the algebra of those operators -- the center of $\mathcal{P}_2$ -- is isomorphic to the center of $\mathcal{D}$, i.e. the only topological operators are the types of particles. For $n\geq 4$ (in our case) additional operators are present -- they correspond to the nonlocal degrees of freedom.
 Examples of topological operators are braiding and fusion operators. As a bonus, this construction provides a way of explicitly calculating the so-called $\mathcal{R}$-matrix i.e. the braiding operator acting on the space $\mathcal{M}$: in the next section we shall see how braiding influences the indices of $F$-operators; writing down the tensors $W^{(k)}(d_1,\ldots,d_n)$ for some set of particles, we can perform explicit index changes corresponding to, say, braiding of $d_i$ around $d_{i+1}$ and see how the $W$-tensors get mapped between each other as a result of that. The tensors form a basis of the protected space  $\mathcal{M}$, so the particular matrix $U_{kk'}$ that we obtain in this calculation is precisely the $\mathcal{R}$-matrix.

Finally let us return to the question of multiplicity of the ground states on the torus. This number is equal to one for the case of the open boundary conditions and 8 for the periodic case, which can be verified by direct computation.  Let us, however, consider a different way of calculating the ground state degeneracy, analogous to the arguments used in Refs.\onlinecite{2007AnPhy.322.1477O, PhysRevB.42.6110}. The general idea is to find a set of operators which all commute with the Hamiltonian, yet do not commute with each other. The topological degeneracy arises from operators with support on non-contractible loops around the torus and the precise value of it can be inferred from their commutation relations.

To this end let us introduce topological charge projectors $K^{R,(C)}_{\sigma}$, which project onto the states such that the total topological charge of all particles in the region bounded by the ribbon $\sigma$ is equal to $R,(C)$ and which together form a resolution of identity (the details of construction of those operators in terms of the ribbons $F^{h,g}_{\sigma}$ may be found in Ref. \onlinecite{2008PhRvB..78k5421B}):
\begin{equation}\label{krcstate} K^{R,(C)}\ket{R,(C');\sigma} = \delta_{RR'}\delta_{(C),(C')}\ket{R',(C');\sigma}, \end{equation}
where $\ket{R',(C');\sigma}$ is a state with a total topological charge in the region bounded by $\sigma$ equal to $R',(C')$. The topological charge projectors commute with the Hamiltonian and commute with one another, and therefore the ground states subspace may be written in a basis in which the basis vectors are eigenstates of the $K^{R,(C)}$'s. Consider now a set of the operators $K_{\sigma}^{R,(C)}$ with $\sigma$ being  the non-contractible loop $\sigma_1$ around one of the holes of the torus, and consider a ground state $\ket{vac;\sigma_1}$ for which $K^{Id,(e)}_{\sigma_1} \ket{vac;\sigma_1} = \ket{vac;\sigma_1}$. The topological charge on the loop $\sigma_1$ may be changed by the application of the operators $F^{ R',(C')}_{\sigma_2}$ creating, winding and annihilating a pair of quasiparticles along the non-contractible loop $\sigma_2$ around \emph{the other} hole of the torus. The action of the operators $K_{\sigma_1}^{R(C)}$ and $F_{\sigma_2}^{R(C)}$ on the ground state $\ket{vac;\sigma_1}$ satisfies:
\begin{gather}[ K^{R,(C)}_{\sigma_1},F^{ R',(C')}_{\sigma_2}  ] \ket{vac;\sigma_1} =\mbox{\ \ \ \ \ \ \ \ \ \ \ \ \ \ \ \ \ \ \ \ \ \ }\nonumber \\
= \left( \delta_{R,R'}\delta_{(C),(C')} - \delta_{R,Id}\delta_{C,(e)}\right)F^{ R',(C')}_{\sigma_2}\ket{vac;\sigma_1}. \label{krcfrc} \end{gather}

Since the operators $F_{\sigma_2}^{R(C)}$ commute with the Hamiltonian, the state $F^{ R',(C')}_{\sigma_2}\ket{vac;\sigma_1}$ is a ground state of the Hamiltonian as well. Eq. (\ref{krcfrc}) guarantees that this state is orthogonal to $\ket{vac;\sigma_1}$. Thus, this procedure creates a subspace of ground states whose dimension is the number of particle types, namely eight.

%
%

We could now, in principle, consider the same sets of operators but with ribbons interchanged, i.e. the topological charge projectors on ribbon $\sigma_2$ and $F^{R',(C')}_{\sigma_1}$ on ribbon $\sigma_1$, corresponding to "measuring or threading the topological charge through hole 2 of the torus". Had those operators been independent of the first two sets, we would have obtained another factor of $8$ for the degeneracy. This is not the case, however, since it is easy to see that the linear vector space of operators spanned by the sets $\left\{ K^{R,(C)}_{\sigma_i} \right\}$ and $\left\{ F^{R',(C')}_{\sigma_i} \right\}$  on \emph{the same} non-contractible ribbon $\sigma_i$ are exactly equal -- both of them are simply different linear combinations of $F^{h,g}_{\sigma}$. Thus the relation (\ref{krcfrc}) with ribbons $\sigma_1$ and $\sigma_2$ interchanged is equivalent to the first one. Hence we conclude there is a topological degeneracy of $8$.

\section{Non-abelian statistics: computing braid matrices\label{statint}}

In this section we will explicitly analyze the quasiparticle braiding statistics in our model. 
As in the previous section, we will neglect the electronic degrees of freedom and focus
entirely on the bosonic model, since the electronic degrees of freedom do not contribute to the
non-abelian part of the statistics -- our main focus here.

We assume that the particles are ordered in the 2D plane and that they are created using ribbon operators which start at the (auxiliary, unphysical) site $x_0$. The indices associated to that site are 'topological'. We want to compute the effect of full braiding (i.e. taking one particle around another and returning to the initial configuration) of two particles on the state of the system, which is equivalent to two exchanges of the particles.

 Let us summarize the results before we explicitly derive them. We find that charges do not accumulate topological phases when winding around one another. The mutual charge-vortex statistics is determined by the entries of the matrix, which is the representation (labeling the charge) of the group element in the conjugacy class labeling the vortex. We show how the entries of the $\mathcal{R}$-matrix can be obtained from this representation. It does not in any way depend on the local degrees of freedom.

In particular, we find that when an abelian A-charge winds around a $\tau$-vortex it accumulates a phase of $\pi$, and when it winds around an $r$-vortex it does not accumulate any topological phase. More interesting is the case of a non-abelian $B$-charge going around a $\tau$-vortex, where the physical state of the system  undergoes a nontrivial transformation by an off-diagonal matrix. Such transformations do not commute, hence we obtain non-abelian statistics.

We will also see that the statistical interaction between the vortices amounts to \emph{flux conjugation} \cite{FA198032}. For exchanges involving the $\tau$-vortices this is also a non-abelian transformation. Finally we derive a general formula which allows the computation also for the case of dyonic particles.

In order to prove all of the claims above we first show how the  indices of individual ribbons creating eigenstates change when the particles are exchanged, in particular we demonstrate that only the 'topological' indices are involved. We then use this result on the physical states described by the $W$-tensors and show how the $\mathcal{R}$-matrix can be obtained. For pedagogical reasons we proceed through a sequence of examples of increasing complexity: first we obtain the phase of $\pi$ for the $A$-charge/$\tau$-vortex winding, then we analyze the $B$-charge case and finally we discuss the general formula for the ribbon exchange which can be used for any physical state in the computation of the $R$-matrix.
Technical details of all of the calculations can be found in Appendix B.

Let us then begin by considering two counterclockwise exchanges of a charge and a vortex ribbon operators, where the vortex stands to the left of the charge initially, as shown in Fig. (\ref{rib3}). The detailed calculation is given in Appendix B. The first exchange produces the following effect:

\begin{equation}\fop{1}{Id}{C}{i}{1}{i'}{1}\fop{2}{R}{e}{1}{j}{1}{j'}= \left( \sum_{k} \bar{\Gamma}_R^{jk}(c_i) \fop{2}{R}{e}{1}{k}{1}{j'}\right)\fop{1}{Id}{C}{i}{1}{i'}{1} \label{braid1}. \end{equation}
The vortex was not affected, thus it does not suffer any change upon braiding with a charge. The state of the charge did get affected and the matrix that governs the change of the 'topological' index of the charge is the representation matrix $\Gamma_R$ associated with the charge. Note that $\Gamma_R$ depends on the 'topological' states of vortex and charge given by the indices $i,j$, and is independent of $i',j'$. The local states $i',j'$ do not get changed during the exchange, neither do they affect the 'topological' indices in any fashion. Computing the second exchange, this time with the charge to the left of the vortex yields a trivial transformation, thus the first exchange is equivalent to the full braiding (the fact that one of the exchanges was equivalent to full braiding is basis dependent).

To elucidate the physical significance of the above result let us consider two explicit examples: the abelian braiding of an $A$-charge and a $\tau$-vortex and the non-abelian braiding of the $B$-charge and the $\tau$-vortex. In line with the discussion in sect.(VI) let us assume that the physical state of the system is given by some $W$-tensor contraction:
\[ \sum_i W^{(1)}_{(i,1),(1,1),\ldots}(\tau, A, \ldots) F_{1,(\tau),Id}^{(i,1)(1,1)}F_{2,(e),A}^{(1,1)(1,1)}\ldots, \]
where we put explicitly that for the $A$-particle there are no running indices since it corresponds to 1-element conjugacy class and 1-dim representation, while for the $\tau$-vortex  $i=1,2$ and all the other summations are implicit. From now on we shall not write $\ldots$ in the tensor contractions corresponding to other particles. They are, however, still present. According to eqs.(\ref{braid1},\ref{appbraid1}) after the braiding the physical state will be given by:
\begin{gather} \sum_i W^{(1)}_{(i,1),(1,1)}(\tau, A)\bar{\Gamma}_A^{11}(c_i) F_{1,(\tau),Id}^{(i,1)(1,1)}F_{2,(e),A}^{(1,1)(1,1)} = \nonumber \\
\label{braid2} =- \sum_i W^{(1)}_{(i,1),(1,1)}(\tau, A)  F_{1,(\tau),Id}^{(i,1)(1,1)}F_{2,(e),A}^{(1,1)(1,1)}, \end{gather}
since $\bar{\Gamma}_A$ is the 1-dim alternating representation that is simply equal to $-1$ on all elements of the $(\tau)$ conjugacy class.
Our result is that the physical state got multiplied by a simple factor of $-1$: this is precisely the abelian phase of $\pi$ the $A$-charge picked up upon encircling the $\tau$-vortex.

We can now perform the similar computation for the non-abelian case of $B$-charge going around a $\tau$-vortex, starting again from a physical state described by some tensor $W^{(1)}$:
\[ \sum_{i,j} W^{(1)}_{(i,1),(1,j),\ldots}(\tau, B, \ldots) F_{1,(\tau),Id}^{(i,1)(1,1)}F_{2,(e),B}^{(1,j)(1,1)}\ldots, \]
where now the summation is also over the index $j=1,2$ since $B$ is a 2-dim representation. Also let us assume that in addition to tensor $W^{(1)}$ there are $W^{(2)},\ldots, W^{(p)}$ which together form a basis of the physical (nonlocal) states.  After the braiding we obtain, again using (\ref{braid1}):
\begin{gather}
\label{braid3}=\sum_{i,k} \tilde{W}_{(i,1),(1,k)}(\tau, B) F_{1,(\tau),Id}^{(i,1)(1,1)}F_{2,(e),B}^{(1,k)(1,1)}, \end{gather}
where $\tilde{W}_{(i,1),(1,k)} = \sum_{j} \bar{\Gamma}_B^{jk}(c_i) W^{(1)}_{(i,1),(1,j)}$. Now we can linearly decompose the $\tilde{W}$ in the basis of $W^{(1)},\ldots,W^{(p)}$:
\begin{equation}\label{braid4} \tilde{W}_{(i,1),(1,k)} = \sum_{q=1}^{p} R_{q,1} W^{(q)}_{(i,1),(1,k)}, \end{equation}

Interestingly, we obtain that upon the braiding the physical state $W^{(1)}$ changed into $\sum_{q=1}^{p} R_{q,1} W^{(q)}$, with the coefficients $R_{q,1}$ dependent on the matrix entries of $\bar{\Gamma}_B$. We denoted the coefficients by $R_{q,1}$ because they form the first column of the $\mathcal{R}$-matrix corresponding to this braiding. Clearly the effect is more complex than a simple abelian phase.

The above examples illustrate how formula (\ref{braid1}) can be used to calculate physical effect of particle braiding, which is just a phase in the abelian case and a non-trivial unitary operation in the non-abelian case.

\begin{figure}[tb]
\centerline{
\includegraphics[width=0.9\columnwidth]{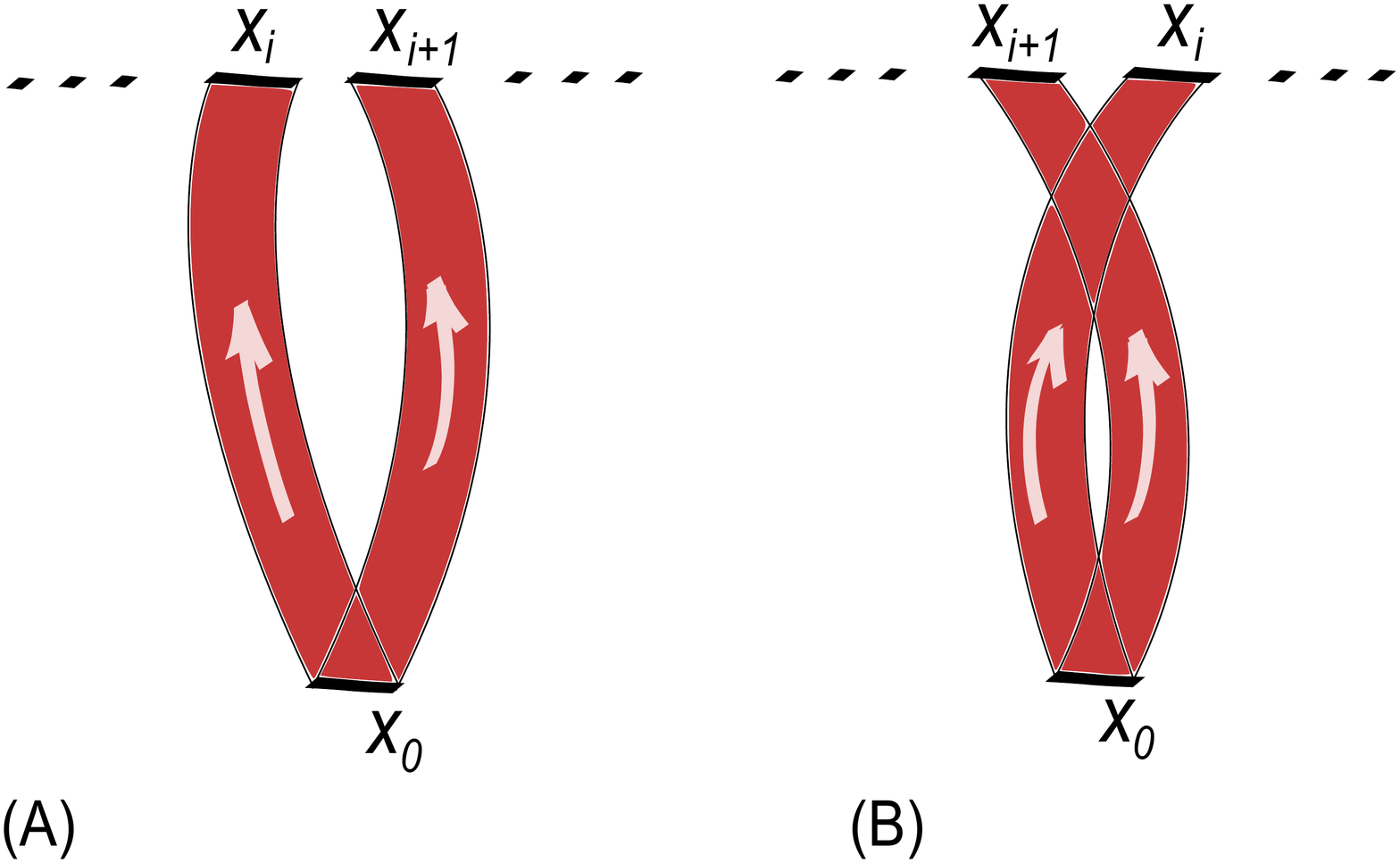}
}
\caption{The exchange of particles at sites $i$ and $i+1$ effected by the $R$ operator. The respective ribbon operators have to be commuted.}
\label{rib3}
\end{figure}

We are now in position to examine a general braiding of two arbitrary excitations of our system, labeled by the pairs $R_1, (C_1)$ and $R_2, (C_2)$, which we have reduced to calculating the effect of an exchange. It can be shown (see Appendix B) that:
\begin{align}&\ \ \fop{1}{R_1}{C_1}{i_1}{j_1}{i_1'}{j_1'} \fop{2}{R_2}{C_2}{i_2}{j_2}{i_2'}{j_2'} =  \nonumber \\
\label{braid5} &= \left(  \sum_{l} \bar{\Gamma}_{R_2}^{j_2l}(\bar{q}_{i_2}c_{i_1}q_k) \fop{2}{R_2}{C_2}{k}{l}{i_2'}{j_2'} \right) \fop{1}{R_1}{C_1}{i_1}{j_1}{i_1'}{j_1'}.  \end{align}

In general (for example in the case of vortex-vortex) also the second exchange is nontrivial, but has the same form, hence we can describe the physical consequences based on result (\ref{braid5}). Note again that local indices are not involved at all, only 'topological' degrees of freedom matter.

Two things happen to the particle $R_2, (C_2)$ upon exchange with $R_1, (C_1)$. The 'flux' degree of freedom $c_{i_2}$ gets conjugated by the other particle to $c_k = \bar{c}_{i_1}c_{i_2}c_{i_1} $. This is the \emph{flux conjugation}. The charge degree of freedom is acted upon by the matrix-representation $\Gamma_{R_2}$ not of $c_{i_1}$ -- which in general does not have to belong to the normalizer $N_{C_2}$ -- but of a related element $\bar{n}_2' = \bar{q}_{i_2}c_{i_1}q_k $, which does. This is expected, since $R_2$ is a representation of the normalizer and not of a full group and hence $c_{i_1}$ has to be 'projected' to the normalizer. This algebraic result is fully consisted with the general considerations in the framework of discrete gauge theories.

Note that the vortex-charge exchange we derived before is a special case. So is the vortex-vortex exchange, which only involves flux conjugation (of both particles, the other one in the second exchange). Conjugations involving elements in the $(\tau)$ conjugacy class are nontrivial, hence the $\tau$-vortices have mutual non-abelian statistics. 

\section{Conclusions and outlook\label{conclusions}}

In this paper, we have constructed exactly soluble lattice models realizing a non-abelian $\nu = 1$ quantum Hall state and a non-abelian 2D fractional
topological insulator. These quantum Hall and topological insulator states have a number of interesting features: they support 
excitations with fractional quantum numbers and non-abelian statistics, they have protected gapless edge modes, and
they have nonvanishing electric Hall and spin Hall conductivities.

An extension of these models to 3D non-abelian fractional topological insulators is also possible. The first step is to
construct a 3D non-abelian bosonic model. This can be accomplished using the 3D generalized toric code 
Hamiltonian with group $\mathbb{D}_3$. This model has two types of excitations: point-like excitations called ``charge'' 
particles and extended string-like excitations called ``vortex loops.'' The charge particles come in two types -- 
$A$-charges and $B$-charges -- and similarly the vortex loops can be either $r$-vortices or $\tau$-vortices. 
In the second step, we couple the 3D bosonic model to electrons in such a way that each electron becomes
bound to an $A$-charge. The resulting composite particle is a spin-$1/2$, charge $e$ fermion. In the third step we 
put the composite fermions into a 3D topological insulator band structure.

As in the 2D case, the resulting insulator supports excitations with non-trivial braiding statistics. These statistics are
inherited from the underlying bosonic model. Several types of braiding operations are possible in this system, including 
winding a charge around a vortex loop or winding one vortex loop through another vortex loop. The braid matrices corresponding 
to these operations can be either abelian or non-abelian, depending on the particles or vortex loops that are involved. 

In addition to braiding statistics, the 3D non-abelian topological insulator also exhibits interesting surface
physics. In particular, this state has protected surface modes that cannot be gapped out without breaking time reversal
or charge conservation symmetry. To see this, note that $q_f/e^* = e/e = 1$, where $q_f$ is the charge of the
fermion and $e^*$ is the minimal charge. Given that this ratio is an odd number, it follows from the flux insertion argument
in Ref. \onlinecite{Levin:2011hq} that this state must have protected surface modes. In the simplest model for the boundary, 
these surface modes are similar to a conventional topological insulator and are 
characterized by the presence of an odd number of Dirac nodes. The only difference from the non-interacting case is that the
underlying fermions are not electrons, but rather are composite particles built out of an electron and an $A$-charge.

If time reversal symmetry is weakly broken at the surface, the boundary of the non-abelian topological insulator can be
gapped. In this case, the surface will exhibit a surface quantum Hall response with a Hall conductivity which is an odd 
multiple of $e^2/2h$. In addition, the surface can accomodate excitations that cannot enter the bulk, since the bulk gap
is larger than the surface gap. It is natural to wonder: do any of these surface excitations carry fractional charge or 
fractional statistics? The answer to this question is ``no'': the only particle-like surface excitations are the composite 
fermion particles which carry charge $e$ and Fermi statistics. 

On the other hand, it is tempting to think that a vortex line that connects two points at the surface could support 
fractional charge at its two ends. Indeed, in the 2D case, we showed that vortex excitations (e.g. the $\tau$-vortex) carry 
charge $e/2$ in the non-abelian $\nu = 1$ quantum Hall state. By similar reasoning one might think that the ends of the 
vortex line could carry fractional charge in the 3D case. However, a key point is that the charge at the end of a vortex line 
is not well-defined or universal in general. The reason is that there is no symmetry or other principle that prohibits a vortex line 
from acquiring a charge polarization (uniform or oscillating) along its entire length. If the vortex line acquires a uniform 
polarization, then this polarization will modify to the charge at the two ends of the vortex line so that this charge is 
non-universal. On the other hand, if the polarization is oscillating, then this polarization will lead to an oscillating 
charge density along the length of the vortex line, so that the charge at the ends is not well-defined.

For similar reasons, we cannot meaningfully discuss the statistics of an individual end of a vortex line, since such 
ends are not truly point particles. However, it \emph{is} meaningful to consider the braiding staistics of an entire vortex line:
for example, we can imagine braiding two vortex lines through one another in the same way that two vortex loops can be 
braided through one another in the 3D bulk. (Similarly, it is also meaningful to discuss the fractional statistics 
associated with braiding a particle excitation around a vortex line).

The most unambiguous manifestation of a fractional topological insulator in 3D is a surface Hall conductivity that is 
not an odd integer multiple of $e^2/2h$. Such a surface Hall conductivity occurs for the $\mathbb{Z}_k$ models from 
Ref. \onlinecite{Levin:2011hq}, but not for the $\mathbb{D}_3$ model discussed here. The crucial difference is that the 
``charge'' excitations in the bosonic $\mathbb{Z}_k$ model carry fractional charge while the charge excitations in the 
bosonic $\mathbb{D}_3$ model are neutral. It is interesting to consider whether a different non-abelian gauge group could
allow for a bosonic model with fractionally charged excitations. The values of possible fractional charges in the 
theory are constrained on one hand by the structure of the fusion rules, and on the other hand by the form of charge 
conserving terms of the type 
we considered in Ref. \onlinecite{Levin:2011hq}, which allow only values of the form $1/m$ where $m$ is the order of some 
generator of the group. It may be possible to find a small non-abelian group for which those two criteria would allow for 
a nontrivial fractional charge in the bosonic system. In such a system one could couple the electromagnetic field directly 
to the bosons and thus excite vortices by inserting electromagnetic flux. It would be interesting to analyze the surface states of
such a system.

\section{Acknowledgements} 
This work was funded by the Israel-US Bi-National Science Foundation and by the Minerva Foundation.
ML acknowledges support from the Alfred P. Sloan foundation. 

\section{Appendix A: Fusion rules}

Following Ref. \onlinecite{Bas} we can introduce shorthand notations for the particles in the model:
\begin{gather*} \mathbbold{1} := (e),Id  \mbox{\ \ \ \ \ } A := (e),A \\
 K^{a/b} := (\tau), Id; \mbox{\ \ }(\tau),A \\
 J^{w/x/y/z}:= (e),B; \mbox{\ \ } (r),Id; \mbox{\ \ } (r),r1; \mbox{\ \ }(r), r2.\end{gather*}
The fusion rules of the theory are now given by:

\begin{gather} A \times A = \mathbbold{1},\\
 A \times K^{a/b} = K^{b/a},\\
 A \times J^{\alpha} = J^{\alpha},\\
 K^{\alpha} \times K^{\alpha} = \mathbbold{1} + J^w + J^x + J^y + J^z,\\
 K^{a} \times K^{b} = A + J^w + J^x + J^y + J^z,\\
 K^{\alpha} \times J^{\beta} =  K^a + K^b,\\
 J^{\alpha} \times J^{\alpha} = \mathbbold{1}+ A+ J^{\alpha},\\
 J^{\alpha} \times J^{\beta} = J^{\gamma} + J^{\delta},\end{gather}
where $\alpha$, $\beta$, $\gamma$, $\delta$ are running indices, and in the last formula they are assumed to be all different.

\section{Appendix B: Statistical interaction}

In this appendix we shall perform some of the computations alluded to in section(VI) more explicitly.

The result in eq.(\ref{braid1}) on the counterclockwise exchange of a charge and a vortex, where vortex stands to the left of the charge initially, is obtained in the following way:
\begin{align}&\fop{1}{Id}{C}{i}{1}{i'}{1}  \fop{2}{R}{e}{1}{j}{1}{j'} = \sum_{n\in N_C}\sum_{g\in\mathbb{D}_3} \bar{\Gamma}_R^{jj'}(g)\fopb{1}{\bar{c}_i}{q_in\bar{q}_{i'}}\fopb{2}{e}{g}=\nonumber \\
\ \  &= \sum_{n\in N_C}\sum_{g\in\mathbb{D}_3} \bar{\Gamma}_R^{jj'}(g)\fopb{2}{e}{\bar{c}_ig}\fopb{1}{\bar{c}_i}{q_in\bar{q}_{i'}} =  \nonumber \\
\ \ &= \sum_{n\in N_C}\sum_{g'\in\mathbb{D}_3} \bar{\Gamma}_R^{jj'}(c_ig')\fopb{2}{e}{g'}\fopb{1}{\bar{c}_i}{q_in\bar{q}_{i'}} =  \nonumber \\
\ \ &= \sum_{k} \bar{\Gamma}_R^{jk}(c_i)\sum_{g'\in\mathbb{D}_3}\bar{\Gamma}_R^{kj'}(g')\fopb{2}{e}{g'}\sum_{n\in N_C}\fopb{1}{\bar{c}_i}{q_in\bar{q}_{i'}} =  \nonumber \\
\label{appbraid1}  &= \left( \sum_{k} \bar{\Gamma}_R^{jk}(c_i) \fop{2}{R}{e}{1}{k}{1}{j'}\right)\fop{1}{Id}{C}{i}{1}{i'}{1} \end{align}
where in the second line we used the commutation relation (\ref{FopCom}), in the third line we relabeled $\bar{c}_ig = g'$ and in the fourth we used the fact that $\Gamma$ are representation matrices. This formula can be then used to compute the effect of charge-vortex braiding on the physical states, both for the abelian and non-abelian case, as has been shown in eqs.(\ref{braid2},\ref{braid3},\ref{braid4}).

Analogously, a fully general result in eq.(\ref{braid5}) can be obtained. Here we assume two ribbons of particles characterized by the pairs $R_1, (C_1)$ and $R_2, (C_2)$ being exchanged.

\begin{align}&\ \ \fop{1}{R_1}{C_1}{i_1}{j_1}{i_1'}{j_1'} \fop{2}{R_2}{C_2}{i_2}{j_2}{i_2'}{j_2'} =  \nonumber \\
& \sum_{n_1\in N_{C_1}} \bar{\Gamma}_{R_1}^{j_1j_1'}(n_1)\fopb{1}{\bar{c}_{i_1}}{q_{i_1}n_1\bar{q}_{i_1'}}  \sum_{n_2\in N_{C_2}} \bar{\Gamma}_{R_2}^{j_2j_2'}(n_2)\fopb{2}{\bar{c}_{i_2}}{q_{i_2}n_2\bar{q}_{i_2'}} = \nonumber \\
& = \sum_{n_2} \bar{\Gamma}_{R_2}^{j_2j_2'}(n_2)\fopb{2}{\bar{c}_{i_1}\bar{c}_{i_2}c_{i_1}}{\bar{c}_{i_1}q_{i_2}n_2\bar{q}_{i_2'}} \sum_{n_1} \bar{\Gamma}_{R_1}^{j_1j_1'}(n_1)\fopb{1}{\bar{c}_{i_1}}{q_{i_1}n_1\bar{q}_{i_1'}} \nonumber \\
& = \sum_{n_2} \bar{\Gamma}_{R_2}^{j_2j_2'}(n_2)\fopb{2}{\bar{c}_k}{\bar{q}_k n_2'n_2\bar{q}_{i_2'}} \fop{1}{R_1}{C_1}{i_1}{j_1}{i_1'}{j_1'} =  \nonumber \\
& = \sum_{l} \bar{\Gamma}_{R_2}^{j_2l}(\bar{n}_2') \sum_{n_2''} \bar{\Gamma}_{R_2}^{lj_2'}(n_2'')\fopb{2}{\bar{c}_k}{\bar{q}_k n_2''\bar{q}_{i_2'}} \fop{1}{R_1}{C_1}{i_1}{j_1}{i_1'}{j_1'} =  \nonumber \\
\label{appbraid5} &= \left(  \sum_{l} \bar{\Gamma}_{R_2}^{j_2l}(\bar{q}_{i_2}c_{i_1}q_k) \fop{2}{R_2}{C_2}{k}{l}{i_2'}{j_2'} \right) \fop{1}{R_1}{C_1}{i_1}{j_1}{i_1'}{j_1'},  \end{align}
where in the third line we commuted the $F$ operators using (\ref{FopCom}). In the fourth line we used the fact that $\bar{c}_{i_1}c_{i_2}c_{i_1} = c_k$ for some $k$. Furthermore we know that on one hand $c_{i_2} = q_{i_2} c_1 \bar{q}_{i_2}$ and on the other $c_k = q_k c_1 \bar{q}_k$, hence we have:
\begin{equation}\label{ccom} (\bar{c}_{i_1}q_{i_2}) c_1 \overline{(\bar{c}_{i_1}q_{i_2})} = q_k c_1 \bar{q}_k, \end{equation}
which shows that $\bar{c}_{i_1}q_{i_2} = q_k n_2'$ for some normalizer element  $n_2' = \bar{q_k}\bar{c}_{i_1}q_{i_2} \in N_{C_2}$.  In the fifth line we put $n_2'n_2 = n_2''$, change the summation to $n_2''$ and use the fact that $\Gamma_{R_2}$ is a representation matrix. We refer to the discussion below eq.(\ref{braid5}) for the physical significance of this result.

Finally, let us show how the nontrivial $\mathcal{R}$-matrix result for the case of $B$-charge in eqs.(\ref{braid3},\ref{braid4}) is derived. Assuming we begin in a physical state described by the tensor $W^{(1)}$, we have
\[ \sum_{i,j} W^{(1)}_{(i,1),(1,j),\ldots}(\tau, B, \ldots) F_{1,(\tau),Id}^{(i,1)(1,1)}F_{2,(e),B}^{(1,j)(1,1)}\ldots, \]
where now the summation is also over the index $j=1,2$ since $B$ is a 2-dim representation. Also let us assume that in addition to tensor $W^{(1)}$ there are $W^{(2)},\ldots, W^{(p)}$ which together form a basis of the topological space.  After the braiding we obtain, again using (\ref{braid1},\ref{appbraid1}):
\begin{gather} \sum_{i,j} W^{(1)}_{(i,1),(1,j)}(\tau, B)\sum_k\bar{\Gamma}_B^{jk}(c_i) F_{1,(\tau),Id}^{(i,1)(1,1)}F_{2,(e),B}^{(1,k)(1,1)}= \nonumber \\
\sum_{i,k} \sum_{j} \bar{\Gamma}_B^{jk}(c_i) W^{(1)}_{(i,1),(1,j)}(\tau, B) F_{1,(\tau),Id}^{(i,1)(1,1)}F_{2,(e),B}^{(1,k)(1,1)}  = \nonumber \\
\label{appbraid3}=\sum_{i,k} \tilde{W}_{(i,1),(1,k)}(\tau, B) F_{1,(\tau),Id}^{(i,1)(1,1)}F_{2,(e),B}^{(1,k)(1,1)}, \end{gather}
where we denoted $\tilde{W}_{(i,1),(1,k)} = \sum_{j} \bar{\Gamma}_B^{jk}(c_i) W^{(1)}_{(i,1),(1,j)}$.

\section{Appendix C: the $A_s^{R,k}$ operators}

The index $k$ enumerates internal degrees of freedom. As such, it is suppressed for $A$-charges. For $B$- charges, it enumerates a local degree of freedom, since
%
\begin{gather} A_{s_0}^{R,k}F_{s_0,s_1}^{R',(1,j),(1,j')}\ket{gs}_{H_k} = \delta_{R,R'}\delta_{k,j}F_{s_0,s_1}^{R',(1,j)(1,j')}\ket{gs}_{H_k}, \nonumber \\
 A_{s_1}^{R,k}F_{s_0,s_1}^{R',(1,j),(1,j')}\ket{gs}_{H_k} = \delta_{R,R'}\delta_{k,1-j'}F_{s_0,s_1}^{R',(1,j)(1,j')}\ket{gs}_{H_k}, \end{gather}
i.e. the operators $A_s^{R,k}$ project onto the states with a local degree of freedom equal to $k$ or $1-k$, depending only on the end of the ribbon they act on. In particular if we create the particles using ribbons anchored at the auxiliary site $x_0$ they will be oblivious to the value of the 'topological' index.

Although the operators $A_s^{B,k}$ commute with all other terms in the Hamiltonian, they do not commute with all local perturbations. Specifically  perturbations of the form $A_{s,\tau r^m}$ act in the following way: $A_s^{B,1}A_{s,\tau r^m} = A_{s,\tau r^m}A_s^{B,2}$. Since we are really interested in the properties that are insensitive to the local perturbations we should put into the Hamiltonian a symmetric superposition:
\begin{equation}\label{assym} A_s^{B} = A_s^{B,1} + A_s^{B,2}. \end{equation}

\bibliography{nonabelian}
\end{document}